\documentclass[aps,pre,twocolumn,showpacs,superscriptaddress,groupedaddress,10pt]{revtex4-1}  

\usepackage{graphicx}  
\usepackage{grffile}
\usepackage{dcolumn}   
\usepackage{bm}        
\usepackage{amssymb}   
\usepackage{amsmath}
\usepackage{amsthm}
\usepackage{mathtools}
\usepackage{xcolor}
\usepackage{tikz}	
\usepackage{ulem}
\usepackage[ansinew]{inputenc}
\usepackage{calrsfs}

\DeclareMathOperator*{\Tr}{Tr}

\begin{document}



\title{Ensemble dependence of the critical behavior of a system
with long range interaction and quenched randomness}

\date{\today}

\author{Nir Schreiber}
\email[]{nir.schreiber@gmail.com}
\affiliation{Department of Mathematics, Bar Ilan University, Ramat Gan, Israel 5290002}
\author{Reuven Cohen}
\affiliation{Department of Mathematics, Bar Ilan University, Ramat Gan, Israel 5290002}
\author{Simi Haber}
\affiliation{Department of Mathematics, Bar Ilan University, Ramat Gan, Israel 5290002}

\begin{abstract}
We propose a hybrid model governed by the Blume-Emery-Griffiths (BEG) Hamiltonian
with a mean-field-like interaction, where the spins are randomly quenched such that some of them are ``pure'' Ising and the others admit the BEG set of states.
It is found, by varying the concentration of the Ising spins, that the model displays	
different phase portraits in concentration-temperature parameter space,
within the canonical and the microcanonical ensembles.
Phenomenological indications that these portraits are rich and rather unusual are provided.
\end{abstract}

\maketitle

\section{Introduction}
Systems with long range interaction (LRI)
\cite{dauxois2002dynamics,Gross2001-es,LyndenBell1968,LyndenBell1999,Thirring1970,cerruti2001clustering,Campa2000,Tamarit2000,barre2005large,Antoni1995}
are usually associated with a pairwise potential of the form $U(r)\sim r^{-\alpha}$,
where $r$ is the distance between two interacting particles
in a $d$-dimensional space and $0\leq \alpha\leq d$ \cite{Gupta2017}.
Suppose, for simplicity, a system of particles, homogeneously distributed in
a hypersphere of radius $R$ and interacting via a LRI potential. In the large $R$ limit,
the energy per particle of the system is dominated by the integral $\int^R r^{d-\alpha-1}dr$,
associated with the total interaction between a particle located in the center of the hypersphere and the other particles.
Since the integral diverges, the total energy of the system is non-extensive,
that is, it does not scale with the volume $V\sim R^d$
\footnote{For $0\leq \alpha<d$ the energy (per particle) diverges as $V^{1-\frac{\alpha}{d}}$. In the special case where $\alpha = d$ it diverges logarithmically with $V$.}.
While the non-extensiveness property can be corrected by properly scaling the
interaction \cite{Kac1963}, a system with LRI may still suffer from non-additivity of the energy.
In other words, such a system with (rescaled) energy $E$, cannot be divided into two subsystems
with energies $E_1,E_2$, where $E = E_1 + E_2 + o(V)$.

A system is expected to have equivalent thermodynamics within
the canonical and the microcanonical ensembles, provided that its energy is additive.
Conversely, non-additivity of the energy may result in peculiar
microcanonical phenomena (that are not observed in the canonical ensemble) such as negative specific heat \cite{Gupta2017}
or the presence of microstates that are inaccessible to the system,
leading to breaking of ergodicity \cite{Mukamel2005}.

The Blume-Emery-Griffiths (BEG) model \cite{Blume1971,Blume_1966,Capel_1966,Blume_1967,Wang1987,Azhari_2022}
has been proposed to explain phase separation
in a mixture of $\rm{He}^3-\rm{He}^4$ atoms. The model
can be naively thought of as describing a classical ``spin-one'' system,
where the spins can take the usual Ising states and additional state where they are equal to zero.
However, the states $\{0,1\}$ practically distinguish between the two types of atoms, while the role of the state $\{-1\}$, additionally assigned to the $\rm{He}^4$ atoms,
is to conceptualize the usual magnetic order parameter.
The model Hamiltonian, depending on the spins configurations, $\{\sigma\}$, may take the general from ${\cal H}(\{\sigma\}) = {\cal H}_I(\{\sigma\}) + \Delta\sum_i \sigma_i^2$
where ${\cal H}_I(\{\sigma\})$ describes the inter spin coupling 
and the other term, where $\Delta$ is the \textit{crystal field} (CF), distinguishes between Ising and zero states.
Typically, for different CF values, the ground state of the model can have either zero or nonzero energy. Suppose the parameters of ${\cal H}_I(\{\sigma\})$ are chosen such that 
in the absence of the CF, the ground state is negative. Then, there is a special value, $\Delta_s$, where for $\Delta<\Delta_s$ the ground state has a complete magnetic ordering and negative energy, while for $\Delta > \Delta_s$ the ground state is totally nonmagnetic with zero energy.
For $\Delta = \Delta_s$ the ground state has zero energy and it is threefold degenerate.
Thus, $\Delta_s$ makes the zero-energy ground state borderline.
It has been shown \cite{Capel_1966} that the model displays first and second
order transitions when $\Delta$ is varied and that there is no phase transition
for $\Delta>\Delta_s$. The associated first and second order critical
lines meet at a tricritical point \cite{Blume1971}.
For reasons that will become clear later, we define a tricritical point, more generally,
to be a point where the type of the transition is changed.

The BEG model may also be a simple example of a model with LRI.
In \cite{Barr2001} the authors considered a BEG model where ${\cal H}_I(\{\sigma\})$ describes a mean-field-like interaction. 
The authors solved the model in the microcanonical ensemble. They have found, employing the canonical solution \cite{Blume1971}, that the model displays a different critical portrait in CF-temperature plane, within the two ensembles. In particular, the canonical and the microcanonical tricritical points, do not coincide. Analyses of the BEG model with mean-field-like interaction 
where the CF is a quenched random variable \cite{jana2016absence,mukherjee2020emergence},
or where an external magnetic field is applied \cite{mukherjee2021phase},
have been recently made. 
Specifically, in \cite{mukherjee2021phase}, a different canonical and microcanonical critical behavior has been  observed.

The main aim of the present paper is to demonstrate inequivalence of the two ensembles
in a rather general fashion, without interfering with the interaction content
of the model. To be more specific, we consider a hybrid
system subject to the mean-field-like BEG Hamiltonian,
where a random concentration of the spins take only the ``up-down'' Ising states.
For those spins, the CF term becomes redundant.
The other spins can additionally occupy the zero state.
Exact canonical and microcanonical solutions to the model,
keeping the parameters of the Hamiltonian fixed,
give rise to different tricritical points in concentration-temperature space.

The rest of the paper is organized as follows. In Sec. \ref{sec:model}
we introduce our model in more detail and present its solution
in the two ensembles. In Sec. \ref{sec:critical properties}
we carefully analyze the phase portraits of the model
together with some related critical properties.
Concluding remarks are given in Sec. \ref{sec:conclusions}.

\section{Model}
\label{sec:model}
Consider a system of $N$ interacting spins governed by the Hamiltonian
\begin{equation}
\label{eq:H}
{\cal H}(\{\sigma\}) = -\frac{J}{2N}\Big(\sum_i \sigma_i\Big)^2 + \Delta\sum_i\sigma_i^2\;,
\end{equation}
where $J>0$ (we take henceforth $J=1$ for simplicity)
is the ferromagnetic coupling constant 
and the normalization factor $N^{-1}$ assures that the total energy is extensive
\cite{Kac1963}. The spins $\sigma_i,\ i=1,2,...,N$ are not homogeneously populated across the lattice.
Strictly speaking, Ising spins $\sigma_i\in \{-1,1\}$, are chosen with probability $p$ and BEG spins having $\sigma_i\in\{-1,0,1\}$ are chosen with probability $1-p$.
We distinguish between strong sites that host Ising spins
and weak sites with BEG spins.
It should be noted that in the case where $p=0$ (homogeneous BEG model), the
Hamiltonian \eqref{eq:H} has $\Delta_s = \frac{1}{2}$.
In the following, we solve the model in the canonical and the microcanonical ensembles.
\subsection{Canonical solution}{\label{sec:sec1}
We employ the standard Gaussian integral representation of
the partition function, $Z$, to write 
\begin{eqnarray}
\label{eq:Z_main_text}
Z =\sqrt{\frac{N\beta}{2\pi}} \int_{-\infty}^\infty dx
e^{-\frac{1}{2}N\beta x^2}\Tr_{\{\sigma\}}e^{\beta x \sum_i \sigma_i-\beta\Delta\sum_i\sigma_i^2}\;,
\end{eqnarray}
where $\beta$ is the inverse temperature, $T$ (in units where Boltzmann's constant, $k_B$, is equal to one). It is shown in Appendix \ref{sec:appA} that,
applying the saddle point approximation to \eqref{eq:Z_main_text} 
and properly averaging over the strong sites,
the free energy density $\beta f = -N^{-1}\ln Z$ can be written (up to terms $o(1)$)  
\begin{equation}
\label{eq:f_canonical_min}
\beta f = \min_x h(x)\;,
\end{equation}
where 
\begin{eqnarray}
\label{eq:f_canonical}
h(x) 
&=& \frac{1}{2}\beta x^2 - p\ln (2e^{-\beta\Delta}\cosh \beta x)\nonumber\\
&-&(1-p)\ln (1+2e^{-\beta\Delta}\cosh \beta x)\;.
\end{eqnarray}

The minimizer of \eqref{eq:f_canonical}, $x_0$, is the order parameter satisfying
\begin{equation}
\label{eq:magnetization}
x_0 = p\tanh(\beta x_0) + (1-p)\frac{2\sinh(\beta x_0)}{e^{\beta\Delta}+2\cosh(\beta x_0)}\;.
\end{equation}

The critical behavior of the model can be detected by
expanding \eqref{eq:f_canonical} in small $x$, yielding
\begin{equation}
\label{eq:f_canonical_small}
f  - f_0= \min_x\left(Ax^2 + Bx^4 + O(x^6) \right)\;,
\end{equation}
where $f_0$ is the high temperature free energy density and 
\begin{eqnarray}
\label{eq:small_f_A}
A & = & \frac{1-\beta p}{2}-\frac{\beta (1-p)}{e^{\beta \Delta}+2}\;,\\
\label{eq:small_f_B}
B & = &
\frac{\beta^3}{12}\left(p-\frac{(1-p) \left(e^{\beta \Delta}-4\right)}{\left(e^{\beta \Delta}+2\right)^2}\right)\;.
\end{eqnarray}
In order for a second order transition to take place, $A$ must
change sign at the critical temperature while $B$ must be positive.
These imply that the critical line, for a fixed $\Delta$, is obtained by setting $A=0$ to give
\begin{equation}
\label{eq:2nd_Ord_can}
2(\beta-1) = (1-\beta p)e^{\beta\Delta}\;.
\end{equation}
The determination of the canonical tricritical point (CTP) requires the simultaneous
vanishing of $A$ and $B$, giving
\begin{equation}
\label{eq:A_can_zero}
e^{\beta\Delta} = 3\beta - 5\;.
\end{equation}

CTPs are limited to a finite interval of CFs.
To see this, we first recall that the $p=0$ homogeneous model
has a CTP in CF-temperature plane, $(\Delta_0,\beta_0)$, where
for $\Delta < \Delta_0$ the transition is of second order \cite{Blume1971}.
The presence of ``Ising intruders'' $(p>0)$ should not change the transition nature.
Second, there is a marginal CF, $\Delta_r$, where for $\Delta>\Delta_r$ \eqref{eq:A_can_zero}
has no solution. At $\Delta_r$, \eqref{eq:A_can_zero} has a unique solution,
determined by equating the derivatives with respect to $\beta$
of both sides of \eqref{eq:A_can_zero}, that is, $\Delta_r$ must solve $e^{\beta\Delta} = 3/\Delta$. This, together with \eqref{eq:A_can_zero}, yields $\Delta_r \simeq 0.489$.
For a fixed CF, taken henceforth to be $\Delta=0.48$,
the solution to \eqref{eq:2nd_Ord_can} and \eqref{eq:A_can_zero}
gives the CTP 
$(p^\ast,\beta^\ast)\simeq (0.0168,3.2624)$. 

It should be noted that substituting $p=0$ in \eqref{eq:2nd_Ord_can}
recovers the second order line of the pure BEG model, satisfying $\beta = \frac{1}{2}e^{\beta\Delta}+1$ \cite{Barr2001}.
Furthermore, in the pure model, the concurrent solution to \eqref{eq:2nd_Ord_can},\eqref{eq:A_can_zero}
for $\Delta,\beta$ produces the CTP $(\Delta_0,\beta_0)=(\frac{1}{3}\ln 4,3)$ \cite{Barr2001}.

\subsection{Microcanonical solution}\label{sec:sec2}
Let $k$ and $n$ be the number of strong and weak spins, respectively, such that $k + n = N$. 
Denoted by $k_{-},k_{+}$, the number of strong spins taking the values ${-1,1}$ and by $n_{-},n_{0},n_{+}$, the number of weak 
spins taking the values ${-1,0,1}$, respectively.
The total energy \eqref{eq:H} can be written 
\begin{eqnarray}
\label{eq:energy}
{\cal E} &=& -\frac{1}{2N}\left(k_{+} - k_{-} + n_{+} - n_{-}\right)^2\nonumber \\
&+& \Delta \left(k_{+} + k_{-} + n_{+} + n_{-}\right)\;,
\end{eqnarray}
and the number of states with energy ${\cal E}$ reads
\begin{eqnarray}
\label{eq:Omega}
\Omega = \binom{k}{k_-,k_+}\binom{n}{n_{-},n_{0},n_{+}}\;.
\end{eqnarray}
Let $\xi_-,\xi_+$ and $\eta_-,\eta_0,\eta_+$ be the fractions of spins in the strong
and in the weak sites, taking the values $-1,1$
and $-1,0,1$, respectively,
satisfying
\begin{eqnarray}
\label{eq:fractions_only}
\xi_- + \xi_+ &=& 1\;,\nonumber\\
\eta_- + \eta_0 + \eta_+ &=& 1\;.
\end{eqnarray}
We then express the spin numbers 
in terms of the fractions and write, to leading order in $N$,
\begin{eqnarray}
\label{eq:spin_numbers}
k_{-} &=& pN\xi_-,\ k_{+} = pN\xi_+\;,\nonumber\\
n_{-} &=& (1-p)N\eta_-,\ n_{+} = (1-p)N\eta_+\;,\nonumber\\
n_{0} &=& (1-p)N\eta_0\;.
\end{eqnarray}
Normalizing \eqref{eq:energy}, i.e., taking $\epsilon = {\cal E}/N$, yields 
\begin{equation}
\label{eq:energy_frac}
\epsilon = -\frac{1}{2}m^2 + \Delta q\;,
\end{equation}
where $m=\frac{1}{N}\sum_i \sigma_i$ and
$q = \frac{1}{N}\sum_i \sigma_i^2$ are the magnetization and quadrupole moment per site,
respectively, which, with the aid of \eqref{eq:spin_numbers}, take the form
\begin{eqnarray}
\label{eq:m_and_q}
m &=& p(\xi_+ - \xi_-) + (1-p)(\eta_+ - \eta_-)\;,\nonumber \\
q & = & p + (1-p)(\eta_+ + \eta_-)\;.
\end{eqnarray}
It is shown in Appendix \ref{sec:appB} that
\begin{equation}
\label{eq:frac_ratios_derivation}
\frac{k_+}{k-k_+}=\frac{n_+}{n-n_0-n_+}\;,
\end{equation}
stating that the entropy has a maximum when the proportion of up and down spins within the strong and the weak regions, is preserved.
Now, plugging \eqref{eq:spin_numbers} into \eqref{eq:frac_ratios_derivation} leads to 
\begin{equation}
\label{eq:ratio}
\xi_+ / \xi_- = \eta_+ / \eta_-\;.
\end{equation}
Eqs. \eqref{eq:fractions_only},\eqref{eq:energy_frac},\eqref{eq:m_and_q} and \eqref{eq:ratio} enable us to express the fractions in terms of $m,p,\Delta,\epsilon$. This gives
\begin{eqnarray}
\label{eq:fractions}
\xi_+ &=& \frac{2 \epsilon+m^2+2\Delta m}{2 \left(2 \epsilon+m^2\right)}\;,\nonumber\\
\xi_- &=& \frac{2 \epsilon+m^2-2\Delta m}{2 \left(2 \epsilon+m^2\right)}\;,\nonumber\\
\eta_+ &=&\frac{\left(2 \epsilon + m^2 + 2\Delta m\right) \left(2 \epsilon+m^2-2\Delta p\right)}{4\Delta (1-p) \left(2 \epsilon+m^2\right)}\;,\nonumber\\
\eta_- &=& \frac{\left(2 \epsilon + m^2 -2\Delta m\right) \left(2 \epsilon+m^2-2\Delta p\right)}{4\Delta (1-p) \left(2 \epsilon+m^2\right)}\;,\nonumber\\
\eta_0 &=& \frac{2\Delta-2 \epsilon-m^2}{2\Delta(1-p)}\;.
\end{eqnarray}

The entropy density, applying the thermodynamic limit to $N^{-1}\ln \Omega$ 
with the aid of \eqref{eq:Omega} and \eqref{eq:spin_numbers}, reads
\begin{equation}
\label{eq:s}
s = 
-\sum_{i\in\{\pm,0\}} \Big(p\xi_i\ln \xi_i + (1-p)\eta_i\ln \eta_i\Big)\;,\\
\end{equation}
where the term $\xi_0\ln \xi_0$ is replaced with zero so that the sum is well defined.
To find the second order critical line
we insert \eqref{eq:fractions} into \eqref{eq:s} and expand $s$ in small $m$,
\begin{equation}
s = s_0 + am^2 + bm^4 + O(m^6)\;,
\end{equation}
where (taking $\tilde\epsilon = \epsilon/\Delta$)
\begin{equation}
\label{eq:s_0}
s_0 = (\tilde\epsilon-1) \ln \left(\frac{1-\tilde\epsilon}{1-p}\right)+(p-\tilde\epsilon) \ln \left(\frac{\tilde\epsilon- p}{1-p}\right)+\tilde\epsilon \ln 2\;,
\end{equation}
is the zero magnetization entropy, and
\begin{eqnarray}
\label{eq:energy_mic_a}
a &=& -\frac{1}{2\tilde\epsilon} +
\frac{1}{2\Delta}\ln \left(\frac{2-2\tilde\epsilon}{\tilde\epsilon-p}\right)\;,\\
\label{eq:energy_mic_b}
b & = & -\frac{1}{8\Delta^2}\left(\frac{1}{\tilde\epsilon-p}+\frac{1}{1-\tilde\epsilon}\right)
+\frac{1}{4\Delta\tilde\epsilon^2}-\frac{1}{12\tilde\epsilon^3}\;.
\end{eqnarray}
Next, we need to make $a$ and $b$ temperature (instead of energy) dependent. Since, in the high temperature phase and at (second order) criticality,
the entropy is maximized by $m=0$; the two coefficients must be nonpositive,
and the microcanonical definition of the temperature $\beta = \partial s/\partial\epsilon$
should be applied to \eqref{eq:s_0}, giving
\begin{equation}
\label{eq:micro_beta}
\tilde\epsilon = \frac{p e^{\beta \Delta}+2}{e^{\beta \Delta}+2}\;.
\end{equation}
Finally, \eqref{eq:micro_beta} is plugged into \eqref{eq:energy_mic_a} and \eqref{eq:energy_mic_b}
and $a$ is set to zero.
The last step recovers \eqref{eq:2nd_Ord_can}.
The (second order) critical energy $\epsilon_c$ and the critical temperature can be related
by combining \eqref{eq:2nd_Ord_can} and \eqref{eq:micro_beta} together at $\beta_c$. This gives
\begin{equation}
\label{eq:epsilon_c}
\epsilon_c = \frac{\Delta}{\beta_c}\;.
\end{equation}
Similar to the canonical solution, where the CTP
has been determined from the simultaneous elimination of the quadratic
and quartic coefficients in the free energy expansion, the coefficients
$a$ and $b$ are set to zero, leading to
\begin{equation}
\label{eq:mic_temp}
e^{\beta \Delta} = -\frac{4}{3} \beta \Delta(\beta \Delta-3)(\beta-1)-2\;,
\end{equation}
and the simultaneous solution to \eqref{eq:2nd_Ord_can} and \eqref{eq:mic_temp} (with $\Delta = 0.48$)
determines the microcanonical tricritical point (MTP) $(\tilde p^\ast,\tilde \beta^\ast) \simeq(0.0170,3.2905)$ 
which is different to the canonical one. 

To find the microcanonical CF, $\tilde\Delta_r$,
above which MTPs do not exist, we apply to \eqref{eq:mic_temp} procedures similar to those employed in finding
the marginal canonical CF, producing $\tilde\Delta_r\simeq 0.482$.
As in the canonical ensemble, the simultaneous solution to \eqref{eq:2nd_Ord_can},\eqref{eq:mic_temp}
for $\beta,\Delta$ in the homogeneous $p=0$ case, recovers previously known results \cite{Barr2001}.

\section{Critical portrait and magnetization}\label{sec:critical properties}

In this section we discuss some of the implications of
our findings from Sec. \ref{sec:model} on some critical properties of the model.
Fig. \ref{fig:PhaseDiag} displays the phase diagram of the model with $\Delta=0.48$
in the canonical ensemble. The ferromagnetic and paramagnetic phases are separated by
the critical portrait, where the second order branch
admits the solution to \eqref{eq:2nd_Ord_can} and the first order branch obeys the simultaneous solution to $f(x_c)=f(0),\ f'(x_c)=0$, where $x_c$ is the nonzero
critical magnetization. 
Apparently, from the inset of the
figure, \eqref{eq:2nd_Ord_can} generates a multivalued curve in the vicinity of the CTP.
For large enough values of $\Delta$, however,
the temperature is a (continuously differentiable) function of the concentration.
In order to find the marginal CF, $\Delta_m$, separating between multivalued curves and functions, it is useful to rewrite
\eqref{eq:2nd_Ord_can} as
\begin{equation}
\label{eq:p(T)}
p(T)=T-2 \left(1-T\right) e^{-\Delta/T}\;,
\end{equation}
and simultaneously solve $p'(T)=0,\ p''(T)=0$
for $\Delta_m$ and the associated temperature, giving $\Delta_m \simeq 0.655$.

\begin{figure}[hbt!]
\begin{center}
\includegraphics[width = 1.\columnwidth]{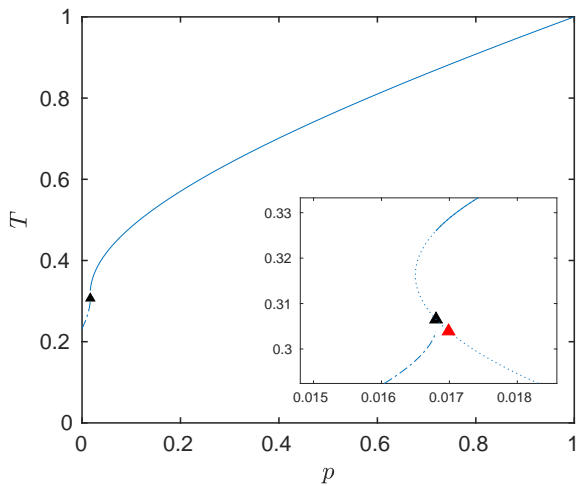}
\caption{Canonical phase diagram in concentration-temperature plane
for $\Delta = 0.48$.
The solid graph corresponds to the second order critical line and the
dashed dotted graph represents the first order line.
The CTP $(0.0168,0.3065)$ is indicated by a black filled symbol (up triangle).
A magnified portion of the diagram, in the
vicinity of the CTP, is displayed in the inset.
The red filled symbol (down triangle) denotes the MTP $(0.0170,0.3039)$.}
\label{fig:PhaseDiag}
\end{center}
\end{figure}

Fig. \ref{fig:can_curves} shows a few curves obeying \eqref{eq:2nd_Ord_can}.
In particular, representatives from the family of multivalued curves are displayed.
At the homogeneous BEG CF $\frac{1}{3}\ln 4$
(and, as turns out, also for $\Delta\gtrsim \frac{1}{3}\ln 4$), the multivalued curve
is made of two branches that are disconnected.
These branches originate from a temperature gap, where $p(T)<0$, that opens up.
\begin{figure}[hbt!]
\begin{center}
\includegraphics[width = 1.\columnwidth]{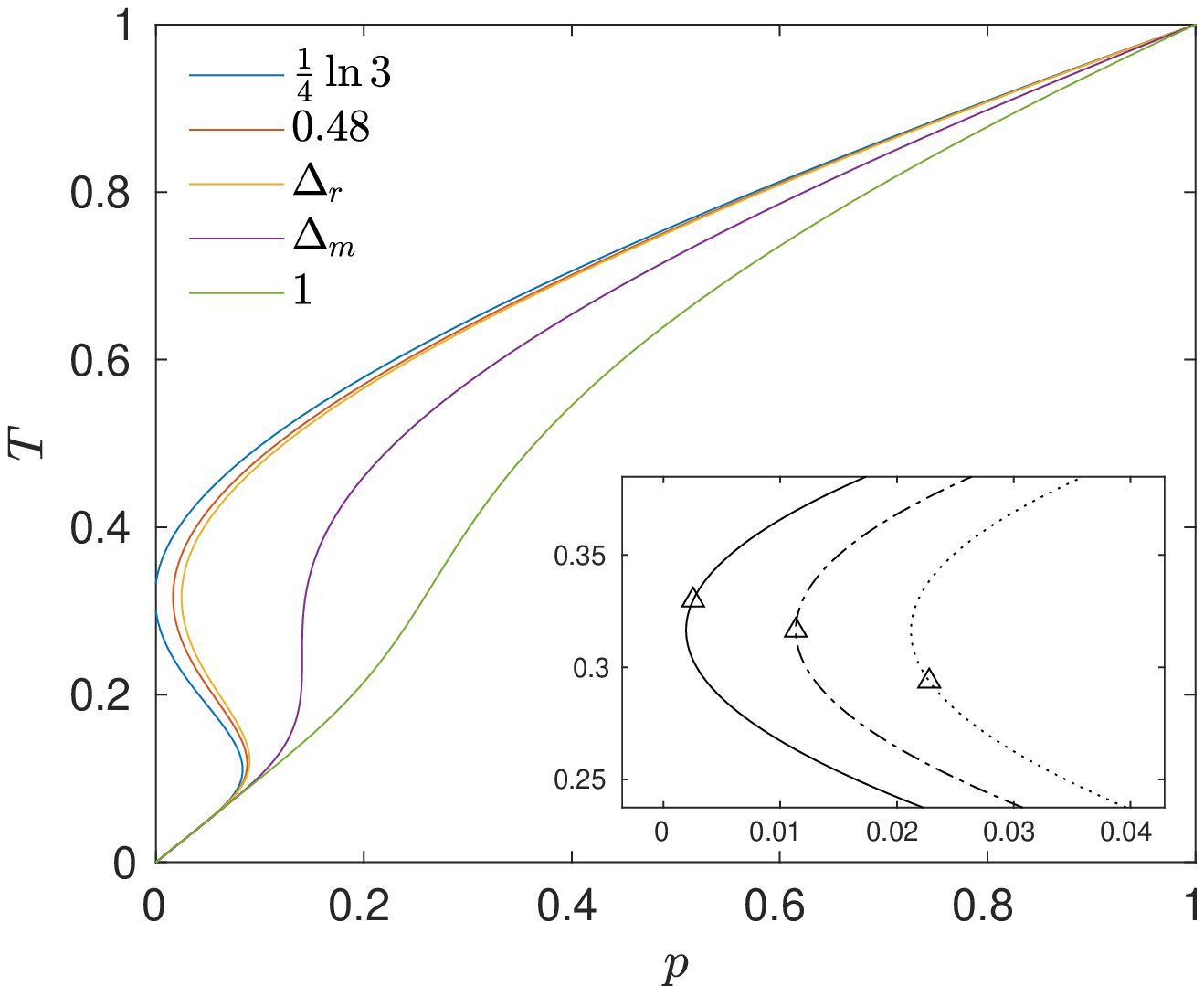}
\caption{Curves obeying \eqref{eq:2nd_Ord_can}, in concentration-temperature plane, for different values of $\Delta$. Graphs from left to right correspond to CFs from top to bottom. Note the small concentrations domain
characterizes the interval $I_{\mathrm{CTP}} = (\frac{1}{3}\ln 4,\Delta_r)$ with $\Delta_r\simeq 0.489$,
where canonical tricritical points survive. Indeed, $\Delta = 0.48$ from Fig. \ref{fig:PhaseDiag} belongs to $I_{\mathrm{CTP}}$.
Inset: Blow up of the regions around the tricritical points (denoted by empty symbols),
where the solid, dashed dotted, and dotted lines
correspond to the CFs $0.465,\Delta_l\simeq 0.475,0.485$, 
respectively.}
\label{fig:can_curves}
\end{center}
\end{figure}
As the inset of Fig. \ref{fig:can_curves} tells,
the position of each CTP on its associated curve,
indicates that the second order line looses continuity at the CTP when the latter is a local maximum of $p(T)$.
This happens at $\Delta_l\simeq 0.475$ which
is part of the simultaneous solution to
\eqref{eq:2nd_Ord_can},\eqref{eq:A_can_zero} and
$2\beta\Delta - 3 = 0$,
where the last equation expresses the condition $p'(T)=0$, in terms of $\beta$.

In summary, a CTP exists for CFs in the interval
$I_{\mathrm{CTP}} = (\frac{1}{4}\ln 3,\Delta_r)$.
Otherwise, for larger values of $\Delta$,
the critical portrait is composed solely from a second order line.
The interval $I_{\mathrm{CTP}}$ can be decomposed into two subintervals.
In the first one, $(\frac{1}{3}\ln 4,\Delta_l)$, the mixed critical portrait
is continuous. It becomes discontinuous in the second one, $(\Delta_l,\Delta_r)$.
Outside $I_{\mathrm{CTP}}$, for $\Delta_r < \Delta <\Delta_m$, the
discontinuity of the critical portrait is expected to survive, even though,
the critical portrait becomes single (second order) typed. 
These discontinuities may result in a second order azeotropy, namely, the simultaneous
exhibition of multiple second order phase transitions
\cite{venaille2009statistical,bouchet2005classification}.
The discontinuous picture is likely to be removed
for $\Delta > \Delta_m$, e.g., for $\Delta = 1$, where the second order line is a function of
the concentration in the interval $(0, 1]$ (see Fig. \ref{fig:can_curves}).
A similar interval composition, with somewhat different boundaries content,
holds in the microcanonical ensemble.

We conclude this section by demonstrating manifestations of ``tricriticality''. first, by means of different behavior of the canonical order parameter \eqref{eq:magnetization} at the critical temperature.
Indeed, as evident from  Fig. \ref{fig:can_mag}, at that temperature, the order parameter
jumps discontinuously for $p<p^\ast$ and it is continuous for $p>p^\ast$.
\begin{figure}[hbt!]
\begin{center}
\includegraphics[width = 1.\columnwidth]{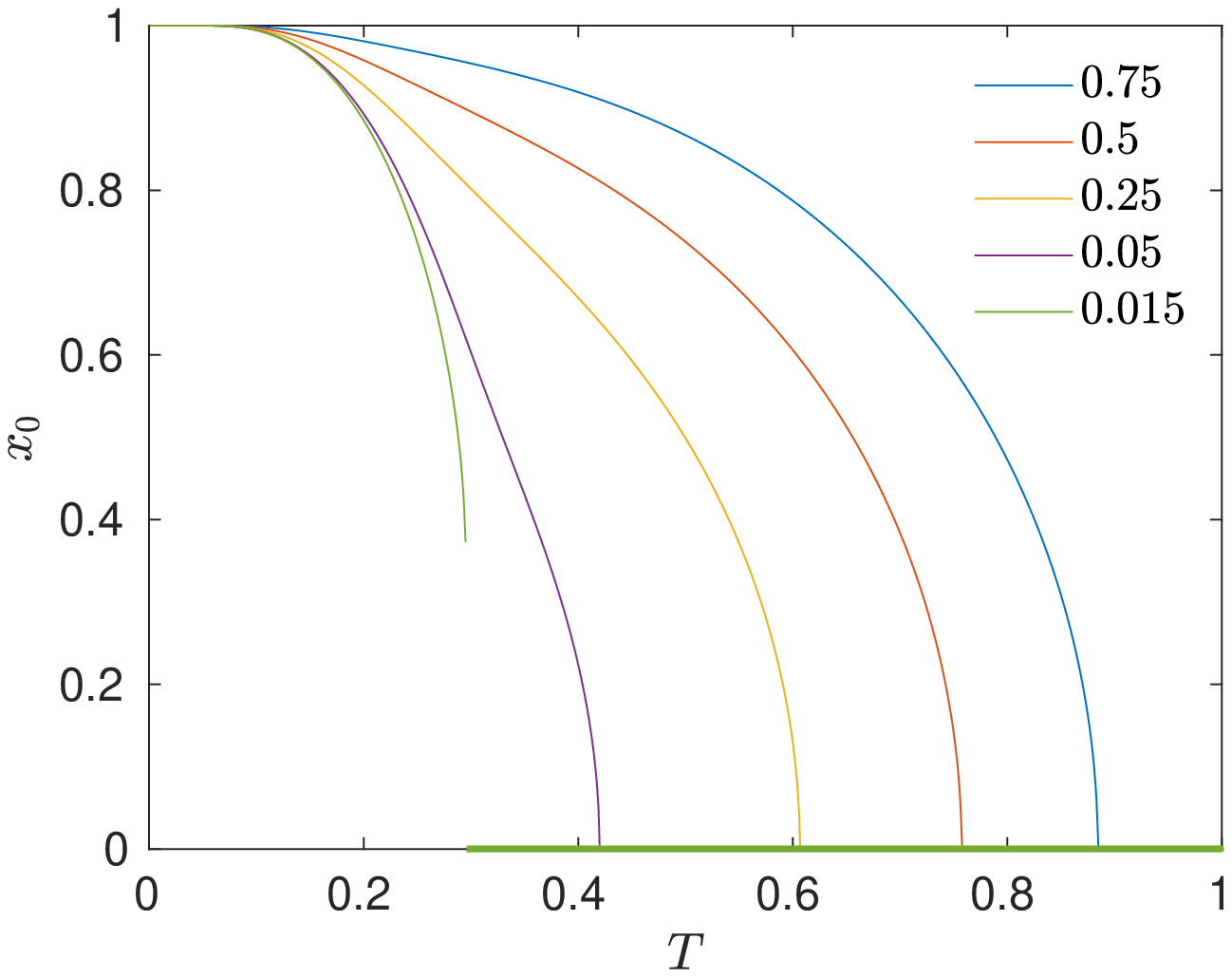}
\caption{Variation of the canonical order parameter satisfying \eqref{eq:magnetization}
with temperature for $\Delta=0.48$ and different concentrations. Graphs from right to left correspond to concentrations from top to bottom. Note the continuous (discontinuous)
behavior at the critical temperature for concentrations above (below)
$p^\ast\simeq 0.0168$. In particular, the jump of the first order magnetization 
(green) is manifested by a composition of a lower thick straight line and an upper line.
}

\label{fig:can_mag}
\end{center}
\end{figure}
Second, Metropolis \cite{metropolis1953equation} Monte Carlo (MC) simulations are performed for a sample of $N=1000$ spins.
The simulated quantities are the total magnetization (per site)
given by the first equation in \eqref{eq:m_and_q}
and the specific energy, proportional to \eqref{eq:H}.
Plots of the latter are presented
in Fig. \ref{fig:s_sim}. The first two charts refer to
the previously used CF $\Delta = 0.48$.
Indeed, the dynamics in these charts discriminates between
first and second order transitions, where in Fig. \ref{fig:s_sim}(a)
the system displays low frequency hops between the coexisting ordered and disordered states
for $p < \tilde{p}^\ast$,
while in Fig. \ref{fig:s_sim}(b) the system hops with high frequency between
the two magnetized states for $p > \tilde{p}^\ast$.
Fig. \ref{fig:s_sim}(c) refers to $\Delta = 1$ where a second order transition is expected
at any concentration. Indeed, small amplitude second order magnetized states,
for a rather small concentration, are evident from this chart.

\begin{figure}[hbt!]
\begin{center}
\includegraphics[width = 1.\columnwidth]{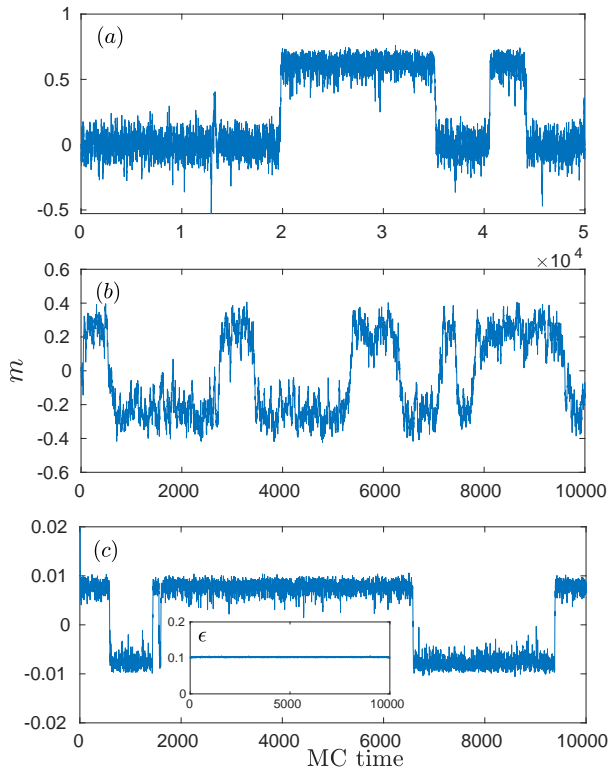}
\caption{Simulated magnetization satisfying the first equation in \eqref{eq:m_and_q} against MC time, for different CFs, concentrations, temperatures and $N=1000$ spins.
The chosen temperatures are in the vicinity of the exact critical temperatures
positioned on the suitable curves in Figs. \ref{fig:PhaseDiag} and \ref{fig:can_curves}.
(a) $\Delta = 0.48$,
$p = 0.0151$, and $T=0.2654$.
(b) $\Delta = 0.48$, $p=0.0385$, and $T=0.3799$.
(c) $\Delta = 1$, $p=0.0998$, and $T=0.0799$.
The specific energy is plotted in the inset.
Its time average (over $50,000$ MC sweeps) and standard deviation are $0.1020$ and $0.0020$,
respectively (c.f. $\epsilon_c = 0.0999$ according to \eqref{eq:epsilon_c}).
}
\label{fig:s_sim}
\end{center}
\end{figure}

\section{Concluding remarks}\label{sec:conclusions}
A hybrid model with mean-field-like LRI and quenched randomness is solved
in the canonical and microcanonical ensembles.
The second order critical lines in concentration-temperature plane are obtained
for the two ensembles.
Indeed, these lines originate from the same solution. However, they eventually terminate in different tricritical points. This may result in different first order critical lines, within the two ensembles, in some interval of small concentrations.

It is found phenomenologically that the model displays rich and rather unusual
phase portraits. Tricritical points are manifested in some interval of CFs.
In some part of that interval, a discontinuity of the second order critical temperature at the tricritical point, is displayed.
A discontinuity of the second order critical temperature is also found for larger CFs outside the interval where the tricritical points exist.
These discontinuities may indicate that multiple simultaneous second order transitions
are exhibited.

Interestingly, the model has no borderline
CF, $\Delta_s$, above which, presumably (as in the pure mode), there is no phase transition
\footnote{Note that the borderline value $\Delta_s=\frac{1}{2}$
of the pure BEG model falls, within the two ensembles, in the domains of CFs where the transition is continuous.}.
Specifically, the model undergoes a second order transition,
with no possible azeotropy, for CFs
outside $I_{\mathrm{CTP}}$ (or the similar microcanonical interval).
This can be easily verified by noting that \eqref{eq:p(T)}
describes a continuous function that becomes monotonic for sufficiently large $\Delta$.
Indeed, for such $\Delta$, by leaving footprints of a second order transition, the simulations (Fig. \ref{fig:s_sim}(c)) may provide another support. 

Special attention should be drawn to the observation that, in the large $\Delta$ regime,
the system may utilize the presence of small concentrations of Ising spins
to eliminate the absence of magnetic ordering characterizing the homogeneous $p=0$ case.
This can be realized by considering \eqref{eq:magnetization} and noting that for 
every large $\Delta$ there is a small $p$ 
such that the order parameter effectively takes the usual Ising form
with $T_c \approx p$. In some sense, a similar phenomenon has been recently detected in another hybrid ($q$-state Potts) model \cite{Schreiber_2022} where, in the large $q$ limit, the system
benefits from the presence of very small concentrations of ``second order'' spins
\cite{duminil2017continuity}; this way avoids a first order transition that would have occurred if those spins where absent
\cite{baxter2016exactly,1611.09877}.


We believe that our approach of randomly quenching spins that respect a subset of states of a known Hamiltonian is rather general and can be applied to other systems with LRI. We expect that some of the findings reported in this paper will be observed in such systems.

\acknowledgments
NS acknowledges support from the Israel Science Foundation (ISF), under Grant No.
$977/17$. 
RC and SH acknowledge support of Bar-Ilan Data Science Institute (DSI) and Israel Council for Higher Education (VATAT). This work was done while SH was visiting the Mathematics Department of Rutgers University-New Brunswick. We thank Professor Gideon Amir for fruitful discussions. The 
constructive critiques of the two anonymous referees who reviewed this work are also acknowledged.

\appendix
\section{Free energy\label{sec:appA}}
In the following, we derive Eqs. \eqref{eq:f_canonical_min} and \eqref{eq:f_canonical}  
for the free energy density.
We start with linearizing the mean-field-like term in the partition function
\begin{equation}
\label{eq:Z}
Z = \Tr_{\{\sigma\}} e^{\frac{\beta}{2N}(\sum_i \sigma_i)^2 -\beta \Delta\sum_i\sigma_i^2}
\end{equation}
by applying the integral identity 
\begin{equation}
\label{eq:gauss}
e^{\frac{\nu ^2}{2\mu}} = \sqrt{\frac{\mu}{2\pi}}\int_{-\infty}^\infty e^{-\frac{1}{2}\mu x^2+\nu x}dx\end{equation}
to \eqref{eq:Z} with $\mu=N\beta$ and $\nu = \beta\sum_i\sigma_i$. This yields
\begin{eqnarray}
\label{eq:Z_app}
Z &=& e^{-N \beta f} = 
\Tr_{\{\sigma\}} e^{\frac{\beta}{2N}(\sum_i\sigma_i)^2-\beta\Delta\sum_i\sigma_i^2}\\
& = & \sqrt{\frac{N\beta}{2\pi}} \int_{-\infty}^\infty dx
e^{-\frac{1}{2}N\beta x^2}\Tr_{\{\sigma\}}e^{\beta x \sum_i \sigma_i-\beta\Delta\sum_i\sigma_i^2}\nonumber\\
&=&  \sqrt{\frac{N\beta}{2\pi}} \int_{-\infty}^\infty dx
e^{-\frac{1}{2}N\beta x^2}
\prod_i\sum_{\sigma_i\in s\cup w} e^{\beta x \sigma_i-\beta\Delta\sigma_i^2}\nonumber\\
& = & \sqrt{\frac{N\beta}{2\pi}} \int_{-\infty}^\infty dx
e^{-\frac{1}{2}N\beta x^2}\nonumber\\
&\times& (2e^{-\beta\Delta}\cosh \beta x)^{N_s}(1+2e^{-\beta\Delta}\cosh \beta x)^{N-N_s}\nonumber\\
&=&\sqrt{\frac{N\beta}{2\pi}} \int_{-\infty}^\infty dx e^{-N h_{N_s}(x)}\nonumber\;,
\end{eqnarray}
where the notation $s\cup w$ refers to the set of ``either strong (Ising) or 
weak (BEG) states''; $N_s\sim \mathrm{Bin}(N,p)$ is the number of strong sites
and
\begin{eqnarray}
h_{N_s}(x) &=& \frac{1}{2}\beta x^2 - \frac{N_s}{N}\ln (2e^{-\beta\Delta}\cosh \beta x)\nonumber\\
&-&\frac{N-N_s}{N}\ln (1+2e^{-\beta\Delta}\cosh \beta x)\;.
\end{eqnarray}
Applying the saddle point approximation to \eqref{eq:Z_app} allows us to write
\begin{equation}
\label{eq:saddle}
N \beta f = N\min_x h_{N_s}(x) + o(N)\;.
\end{equation}
Now, for large $N$ and $N_s$ the Binomial distribution approaches
a normal distribution with the same mean and variance, i.e., $N_s$ obeys
\begin{eqnarray}
& &Pr(N_s = z)\approx \\
& &\frac{1}{\sqrt{2\pi Np(1-p)}}\exp{\left(-\frac{(z-Np)^2}{2Np(1-p)}\right)},\ z\in\mathbb{N}\;.
\nonumber
\end{eqnarray}
This implies that typically  
\begin{equation}
\label{eq:large_Ns}
N_s-\langle N_s\rangle \sim \sqrt N
\end{equation}
and hence
\begin{equation}
\label{eq:h_diff}
N h_{N_s}(x) - N h(x) = o(N)\;,
\end{equation}
where 
\begin{eqnarray}
\label{eq:min_h(x)}
h(x) &=& \langle h_{N_s}(x)\rangle = \frac{1}{2}\beta x^2 - p\ln (2e^{-\beta\Delta}\cosh \beta x)\nonumber\\
&-&(1-p)\ln (1+2e^{-\beta\Delta}\cosh \beta x)\;.
\end{eqnarray}
Combining now \eqref{eq:saddle} and \eqref{eq:h_diff} together leads to
\begin{eqnarray}
\label{eq:f_diff}
 N\beta f - N\beta \langle f\rangle  &\sim & N 
\min_x h_{N_s}(x)-N\langle\min_x h_{N_s}(x)\rangle\nonumber\\
 &=& N\min_x h_{N_s}(x)-N\min_x h(x) = o(N)\;.\nonumber\\
\end{eqnarray}
In other words, it is sufficient to average over the leading order term of the RHS of \eqref{eq:saddle}
in order that the free energy typically deviates from its sample average in an amount of
the same order of magnitude as in \eqref{eq:h_diff}.
Finally, we conclude from \eqref{eq:f_diff} that
\begin{equation}
\beta f = \beta \langle f\rangle + o(1) = \min_x h(x) + o(1)\;.
\end{equation}

\section{Fixed proportion of strong and weak up and down spins\label{sec:appB}}
In the microcanonical ensemble, one fixes the energy and finds the most probable macroscopic state, i.e., the one with the highest entropy. This state corresponds to a maximum number of microscopic configurations.
We derive a necessary condition, involving several counting variables (spin numbers) 
associated with these configurations, for establishing the most probable macroscopic state.
To this end we consider 
the entropy where the latter is expressed in terms of the counting variables $k_-, k_+, n_-, n_0, n_+$
introduced in the main text while keeping the total energy fixed.

We start with introducing the total number of up spins, $t$,
to write the predetermined energy in the form 
\begin{equation}
\label{eq:energy_constraint}
{\cal E} = -(2t + n_0-N)^2/2N + \Delta (N-n_0)\;.
\end{equation}
The entropy can then be written 
\begin{eqnarray}
\label{eq:S_max_k+}
S&=&\ln\left(\binom{k}{k_+}\binom{n-n_0}{t-k_+}\right)\\
&+& S_0 + \lambda \left({\cal E}+ (2t+n_0-N)^2/2N-\Delta(N-n_0)\right)\nonumber\\
&=&\ln\left(\frac{k!(n-n_0)!}{k_+!(k-k_+)!(t-k_+)!(n-n_0-t+k_+)!}\right)
\nonumber\\
&+& S_0 + \lambda \left({\cal E}+ (2t+n_0-N)^2/2N-\Delta(N-n_0)\right)\nonumber\;,
\end{eqnarray} 
where $S_0 = \ln \binom{n}{n_0}$
and $\lambda$ is a Lagrange multiplier
assuring that the entropy is maximized subject to the constraint 
\eqref{eq:energy_constraint}. Note that since $k_+ + k_- = k,\ n_+ + n_- + n_0=n$, 
where $k$ and $n$ are fixed, $S$ depends only on the three variables, $k_+,t,n_0$.
Applying Stirling's approximation to \eqref{eq:S_max_k+}
and setting the derivative with respect to $k_+$ to zero gives
\footnote{It should be noted that the optimization of \eqref{eq:S_max_k+}
can be performed with respect to any of the four up and down spin numbers},
ignoring terms next to leading order,
\begin{eqnarray}
\label{eq:ln_ratio_strong_weak}
\ln(k_+) &+& \ln(n-n_0-t+k_+)\nonumber\\
&=&\ln(t-k_+)+\ln(k-k_+)\;,
\end{eqnarray} 
or, using $t = k_+ + n_+$,
\begin{equation}
\label{eq:ratio_strong_weak}
\frac{k_+}{k-k_+}=\frac{n_+}{n-n_0-n_+}\;.
\end{equation}

To fully optimize \eqref{eq:S_max_k+} one can observe that the first term
in \eqref{eq:energy_constraint} is simply the total magnetization $M$, i.e.,
\begin{equation}
\label{eq:M_constraint}
M  = 2t + n_0-N\;,
\end{equation}
and write \eqref{eq:S_max_k+} as 
\begin{equation}
\label{eq:S_M_n0}
S = \tilde S(M,n_0) + \lambda\left({\cal E} -\varphi(M,n_0)\right)\;,
\end{equation}
where $\tilde S(M,n_0$) is the combinatorial term containing the information in \eqref{eq:ln_ratio_strong_weak} and \eqref{eq:M_constraint},
and 
\begin{equation}
\label{eq:S_M_n0_constraint}
\varphi(M,n_0) = - M^2/2N + \Delta(N-n_0)\;.
\end{equation}
One may then properly optimize \eqref{eq:S_M_n0}, provided \eqref{eq:S_M_n0_constraint},  with respect to $M,n_0$.

In the main text, the (normalized) strong and weak spin numbers,
where the latter are replaced with their expected values,
are functions of the magnetization and energy densities,
$m = M/N$ and $\epsilon = {\cal E}/N$, respectively.
This allows to realize the entropy density as $s(m,\epsilon)$,
thus, is an alternative to the Lagrange multiplier formulation presented here.
Finally, the equivalent treatment to the optimization of \eqref{eq:S_M_n0}
would be to optimize $s(m,\epsilon)$ with respect to $m$.

For the sake of clarity we state that we did not
perform the full optimizations described in this appendix,
simply because we did not need them in our microcanonical analysis. 


%


\end{document}